\documentstyle[epsf,aps]{revtex}
\renewcommand{\thefootnote}{\fnsymbol{footnote}}
\begin{document}
\newcommand{\be}{\begin{eqnarray}}
\newcommand{\dlq}{\lq\lq}
\newcommand{\ee}{\end{eqnarray}}
\newcommand{\ben}{\begin{eqnarray*}}
\newcommand{\een}{\end{eqnarray*}}
\renewcommand{\baselinestretch}{1.0}
\newcommand{\as}{\alpha_s}
\def\eq#1{{Eq.~(\ref{#1})}}
\begin{flushright}
BNL--NT--00/18 \\
TAUP--2637--2000\\
July 17, 2000
\end{flushright}
\vspace*{1cm} 
\setcounter{footnote}{1}
\begin{center}
{\Large\bf QCD Instantons and the Soft Pomeron}
\\[1cm]
Dmitri \ Kharzeev, $^{1,2}$ Yuri V.\ Kovchegov, $^{1,3}$ Eugene
Levin $^{1,3}$ \\ ~~ \\
{\it $^1$ Physics Department, Brookhaven National Laboratory} \\ 
{\it Upton, NY 11973, USA} \\ ~~ \\
{\it $^2$ RIKEN--BNL Research Center, Brookhaven National Laboratory} \\ 
{\it Upton, NY 11973, USA } \\ ~~ \\
{\it $^3$ HEP Department, School of Physics and Astronomy } \\ 
{\it Tel Aviv University, Tel Aviv 69978, Israel } \\ ~~ \\ ~~ \\
\end{center}
\begin{abstract}

We study the r{\^o}le of semi--classical QCD vacuum solutions in high
energy scattering by considering the instanton contribution to
hadronic cross sections. We propose a new type of instanton--induced
interactions (``instanton ladder'') that leads to the rising with
energy cross section $\sigma \sim s^{\Delta}$ of Regge type (the
Pomeron). We argue that this interaction may be responsible for the
structure of the soft Pomeron.  The intercept $\Delta > 0$ is
calculated. It has a non--analytic dependence on the strong coupling
constant, allowing a non-singular continuation into the
non-perturbative region. We derive the Pomeron trajectory, which
appears to be approximately linear in some range of (negative)
momentum transfer $t$, but exhibits a curvature at small $t$. Possible
r{\^o}le of instantons in multiparticle production is also discussed.

\end{abstract}
\renewcommand{\thefootnote}{\arabic{footnote}}
\setcounter{footnote}{0}

\section{Introduction}

QCD vacuum is known to possess a rich structure. A striking example is
provided by the existence of (Euclidean) classical solutions --
instantons \cite{bpst}, which are responsible for non-trivial
topological properties of the theory \cite{th,cdg,jr}. While the
influence of instantons on the properties of hadrons and their
interactions at low energies have been extensively studied
\cite{cdg,es,dd}  (for a review, see \cite{et}), the r{\^o}le of
topological effects in high energy collisions is still an open and
fascinating problem \cite{bj}.

In electroweak theory, significant interest was excited by the
possibility of baryon number non--conservation caused by instantons at
the collision energy above $10$ TeV \cite{ring,mvv,tin,son,mu,rs}. In
QCD, the instanton contribution to the structure functions of
deep--inelastic scattering was studied in Refs. \cite{bb,rslat}, while
the cross section for gluon--gluon scattering mediated by instantons
was considered in Refs. \cite{mu,br,bb2,zakh}.

Recently, it has been proposed \cite{kl} that the existence of
semi--classical solutions in QCD can have important consequences also
for hadron scattering at high energies, inducing a non--perturbative,
and possibly dominant, contribution to the scattering amplitude. This
approach was shown to lead to the ``soft'' pomeron with the intercept
$\alpha_P - 1 \simeq 0.1$ which has a non--trivial dependence on the
strong coupling and the numbers of colors and flavors \footnote{
Analogous non--perturbative behavior was also established in the
scattering of color dipoles at low energy \cite{fk}.}.  The method
used in that work was based on low energy theorems of broken scale
invariance \cite{nsvz}, and did not assume any specific form of the
semi--classical solutions. This approach was based on the low energy
theorems \cite{nsvz} for the trace of the QCD energy-momentum tensor
taken in the chiral limit of massless quarks
\be\label{emt}
\theta_\mu^\mu \, = \, - \frac{b g^2}{32 \pi^2} \, F^{a \alpha \beta} \, 
F^a_{\alpha \beta},
\ee
where $b = (11 N_c - 2 N_f)/3$ and $F^{a \alpha \beta}$ is the gluonic
field strength tensor. The low energy theorems derived in \cite{nsvz}
state that
\be\label{lowen}
i \, \int \, d^4 x \, (0| T \theta_\mu^\mu (x) \theta_\nu^\nu (0) |0)
\, = \, - 4 (0| \theta_\mu^\mu (0) |0).
\ee
Comparing Eqs. (\ref{emt}) and (\ref{lowen}) one can conclude that the
low energy theorem (\ref{lowen}) can only be satisfied by a strong
gluonic field \footnote{Unfortunately the applications of low energy
theorems of \eq{lowen} depends very much on the way one subtracts the
perturbative contribution to the correlation functions
\cite{nsvz}. Lattice simulations have also encountered similar
problems. We thank Al Mueller for clarifying this point to us.}
\be\label{qcl}
A_\mu \, \sim \, \frac{1}{g}.
\ee
The strong field of \eq{qcl} is usually associated with the
quasi-classical solutions of QCD. This quasi-classical field has been
employed in \cite{kl} to provide gluonic interactions, which, after
being iterated in the t-channel gave rise to a pomeron-like behavior
of the corresponding cross sections.

Here we would like to develop the idea proposed in \cite{kl} and
explore the effects of quasi--classical fields on high energy
scattering. Therefore, for the purpose of this paper we will consider
a particular type of this classical field: we will assume that the
field is given by the instanton solution
\cite{bpst}.  We will employ instanton fields, which would allow us to 
quantify the assumption of the quasi--classical fields being
responsible for the pomeron--like behavior of the cross
sections. Nevertheless we can not rule out the possibility of some
other quasi--classical gluonic fields giving a significant
contribution to the processes considered below.

The instanton calculations are better justified in the electroweak
theory, where the coupling is small and one can hope that quantum
corrections to the classical instanton solution are small.  Therefore
the calculation that will be presented below can also be repeated for
the electroweak interactions, where it would possibly also give a
cross section that rises with energy \cite{fut1} \footnote{We thank
Larry McLerran for bringing our attention to this problem.}.

Shuryak \cite{es1} has argued that the explicit use of instantons
could provide a useful way to interpret and extend the results of
\cite{kl}. Very recently, Shuryak and Zahed \cite{sz} analyzed
Euclidean scattering induced by the instantons, and analytically
continued their results to Minkowski space using the correspondence
between Euclidean angle and Minkowski rapidity
\cite{meggiolaro}. Their result is a constant cross section, which
does not rise with energy.

The purpose of the present paper is to investigate the effect of the
instantons on the dynamics of gluon ladders in high energy
scattering. Our approach here is complementary to the one taken in
Ref. \cite{kl}; even though it is hindered by significant numerical
uncertainties, it provides a deeper insight into the mechanism of
non--perturbative effects in high energy scattering. As will become
clear later, our treatment of the problem is different from
Ref.\cite{sz}; while the authors of that paper consider a purely
classical contribution to the scattering amplitude, we attempt at
evaluating the leading quantum terms arising in perturbation theory
built in the background of classical instanton fields.

\begin{figure}
\begin{center}
\epsfxsize=5cm
\leavevmode
\hbox{ \epsffile{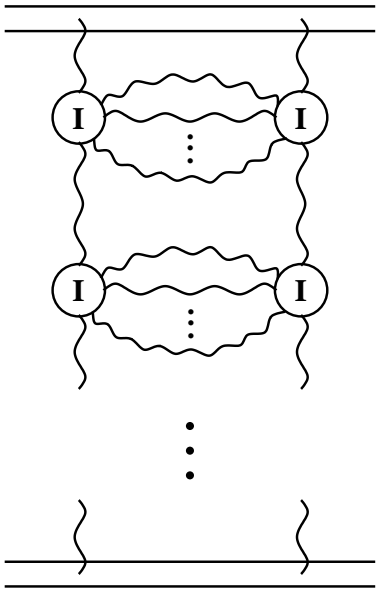}}
\end{center}
\caption{The structure of soft pomeron as conjectured in the paper.}
\label{pom}
\end{figure}

The proposed mechanism of the pomeron is illustrated in
Fig. \ref{pom}. The pomeron is being constructed by resummation of the
ladder diagrams, similarly to the well-known BFKL pomeron
\cite{EAK,Yay}. The emission vertices of our pomeron ladder are given by 
the multiple gluon interaction vertices generated by the instantons
\cite{ring}. The vertices connect to two gluons in the t-channel and 
produce several gluons in the t-channel. We sum over all possible
numbers of the produced t-channel gluons in each vertex
\cite{ring,mvv}. The t-channel lines in the ladder of Fig. \ref{pom} 
will be taken as the usual gluon lines for most of our paper. In
principle one should include the virtual corrections in the t-channel,
which may lead to reggeization of the gluons \cite{EAK,Yay}. We will
present an estimate of these virtual corrections. However the exact
calculation and the answer to the question of whether gluon
reggeization happens in our case will be addressed elsewhere
\cite{fut2}. For most of this paper we will work in pure gluodynamics 
with no quarks. However, at the end of the paper we will present an
estimate of the effect of inclusion of light quarks.

The multiple gluon vertices generated by instantons in the pomeron of
Fig. \ref{pom} will be calculated in Sect. III at the classical
level. The vertices include a suppression factor of $\exp \left( -
\frac{2 \pi}{\as}\right)$, which is due to the classical instanton action. If 
the coupling constant is sufficiently small ($\as \ll 1$), then this
suppression factor is also small
\be
\exp \left( - \frac{2 \pi}{\as}\right) \ll 1 
\ee
and can be used as a parameter justifying the perturbative expansion
of Fig. \ref{pom}. The scale of the running coupling in this case is
set by the typical size of the instantons, so that $\as = \as
(1/\rho_0)$. We assume that $\as (1/\rho_0) \ll 1$, which is only
marginally satisfied in QCD.  Our ladder is resumming the leading
logarithms of center of mass energy $s$, which in this case implies
resummation of all powers of the parameter
\be
e^{- \frac{4 \pi}{\as}} \, \ln s \, \sim \, 1.
\ee
Note that the dependence of our small parameter on $\as$ is
non-analytical, which will introduce the non-analytical dependence on
$\as$ in the calculated value of the pomeron intercept.

As energy in each rung of the ladder increases, quantum corrections to
the tree level calculation of the vertices become important. The
quantum corrections are known to modify the energy dependence of the
$2 \rightarrow n$ transition vertices \cite{ring,mvv,mu,rs}.  The
energy dependence of the $2 \rightarrow n$ total (summed over $n$)
cross section is given by the following formula
\be\label{hg}
\sigma_{2} (s) \, \sim \, \exp \left[ \frac{1}{g^2} \, F \left( 
\frac{\sqrt{s}}{E_{sph}}\right) \right]. 
\ee
Here $s$ is the center of mass energy of the system, $E_{sph}$ is the
sphaleron energy. $F$ is known only for the small values of its
argument \cite{rs}.  The lowest order term in $F$ is given by the tree
level calculations, and the quantum corrections provide higher order
terms in the expansion of $F$ \cite{ring,mvv,mu,rs}.  It is also known
that for a large number of gluon legs (large $n$) the value of the
saddle point is shifted. Thus for large number of produced gluons the
vertices may not be exactly the same instanton--induced vertices as
for small $n$. This only substantiates our argument that the instanton
field may not be an exact mechanism responsible for the multiple gluon
interactions in the ladder of Fig. \ref{pom}. There are many problems
related to the determination of the function $F (\sqrt{s}/E_{sph})$
\cite{rs} and we are not going to review all of them here. We will simply
point out that the growth with energy of the cross section of \eq{hg}
should eventually stop due to the unitarity constraint \cite{zakh}
\be\label{uni}
\sigma \, < \, \frac{const}{s} \, \exp \left( - const \, \frac{2 \pi}{\as} 
\right).
\ee
Moreover, it is possible that the discussed $2 \rightarrow n$ cross
section will fall off at the energies of the two incoming gluons much
higher than the sphaleron energy, since the instanton effects may not
be important in that kinematic region. Assuming that this is true
throughout the paper we will approximate the multi gluon interactions
by the tree level calculation for the energies below the sphaleron
energy and we will just put it to be zero for $\sqrt{s} \, > \,
E_{sph}$. This approximation is rather crude and will be improved in
the subsequent work \cite{fut2}, where several different behaviors of
the $ 2 \rightarrow n$ scattering cross section will be considered. In
particular we will consider the case when the unitarity limit of
\eq{uni} has been reached.

\begin{figure}
\begin{center}
\epsfxsize=8cm
\leavevmode
\hbox{ \epsffile{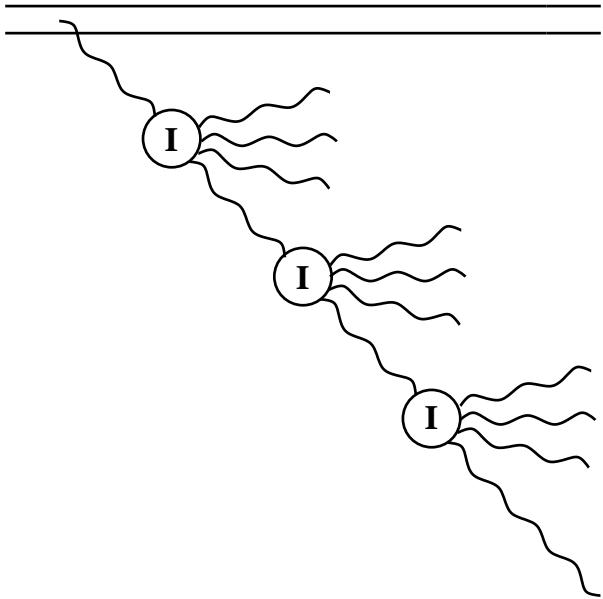}}
\end{center}
\caption{Space--time picture of soft pomeron.}
\label{stp}
\end{figure}

Let us discuss the (Minkowskian) space--time picture of high energy
scattering amplitude that we are going to evaluate.  Since at high
energies the scattering occurs over large longitudinal distances, in
order to construct the amplitude one may need to have several
instantons involved in the process (see Fig.
\ref{pom}). At first glance, this seems to contradict the
assumption of the dilute instanton gas \cite{shur} that we are going
to use\footnote{A computation in a self--consistent instanton vacuum
model would be a very interesting extension of this work.}.  The
resolution of this apparent problem is in the very fact that a pomeron
exchange is not an instantaneous interaction. Pomeron exchange is best
represented in the wave function picture of high energy interactions
\cite{wf}.  In that formalism in order to exchange a pomeron the
colliding hadrons have to develop large multi--gluon fluctuations,
which, in turn, interact with each other, which corresponds to the
pomeron exchange in the usual t--channel language.

The space--time picture of our soft pomeron constructed in the spirit
of the wave function formalism is shown in Fig.  \ref{stp}. Single
pomeron exchange corresponds to a single gluon, which, after being
emitted off the original hadron, propagates through space, inducing a
chain of instanton transitions between different topological vacua and
producing more gluons in each transition. The coherence length of the
small-$x$ gluon in Fig. \ref{stp} is large, $\tau \sim k_+ /
k_\perp^2$, where $k$ is the gluon's momentum. Thus a high energy
gluon carrying a large ``plus'' component of momentum can propagate
over large longitudinal distances inducing several instanton
transitions. Even in the dilute instanton gas approximation
\cite{shur} the gluon should therefore be able to generate many instanton
transitions while traveling over large distances in the longitudinal
direction at the given value of the impact parameter.

If the process is viewed in the rest frame of one of the hadrons, then
after several instanton transitions, the propagating gluon would
interact with the target hadron at rest. In the center of mass frame
the propagating gluon would find another gluon coming from the second
hadron and would interact with it through an instanton transition,
producing more gluons.

Multiple pomeron exchanges in this picture would correspond to the
case when the gluons produced in the interactions of Fig. \ref{stp} in
turn start propagating through the instanton gas and interacting
\cite{me}.  However, a study of that interesting scenario is beyond
the scope of our paper. Here we will just argue that these multiple
pomeron exchanges are suppressed compared to the single pomeron
exchange contribution, for the reasons described below. Of course they
will become important as the center of mass energy of the collision
becomes very high.

The paper is organized as follows: after discussing the general issues
related to instantons in Sect. II we will proceed to calculating the
intercept of the pomeron of Fig. \ref{pom} in Sect. III. We will then
calculate the slope and trajectory of the pomeron in Sect. IV and
conclude by summarizing our results in Sect. V.

\section{General Issues}

In Minkowski space, instantons correspond to tunneling transitions
between different topological vacua \cite{th,cdg,jr}. The instanton
calculations therefore make sense only at energies much smaller than
the energy of the potential barrier separating the vacua
(``sphaleron'' energy $E_{sph}$).

The BPST instanton solution in the regular gauge is given by
\cite{bpst}
\be 
A_\mu^a (x) \, = \, \frac{2\ {\eta_{a \mu \nu}} x_\nu}{g \, (x^2 +
\rho^2)},
\ee
and the corresponding field strength is 
\be
(G^a_{\mu\nu})^2 = \, \frac{192\ \rho^4}{g^2 \, (x^2 + \rho^2)^4}.
\ee
The energy along the instanton path can be defined as 
\be
E(\tau) = {1 \over 8} \ \int d^3x \ (G^a_{\mu\nu})^2,
\ee
and its maximum at $\tau=0$ corresponds to the sphaleron energy (see
\cite{lms}):
\be
E_{sph} = E(\tau=0) = \frac{ 3\ \pi^2}{\rho}\ {1 \over g^2} = \frac{3
\pi}{4 \, \alpha_s \, \rho}.
\ee
To estimate this quantity, we take $\rho \simeq (700\ {\rm
MeV})^{-1}$, and $\as (1/\rho) = g^2/4\pi \simeq 0.7$; we get $E_{sph}
\simeq 2.4 \div 2.6 \ {\rm GeV}$.

At first glance, this value of $E_{sph}$ suggests that the
perturbative treatment about the instanton solution is not applicable
for hadron scattering at high energies $\sqrt{s} >> E_{sph}$. However,
this conclusion is premature. Indeed, the space--time picture of high
energy scattering discussed in the Introduction implies that the
scattering is described by the ladder--type diagrams, where the number
of rungs is proportional to rapidity $y \sim ln s$, and the energy in
each rung $E_{rung}$ is independent of the total collision energy
($E_{rung}$ is typically on the order of a few GeV \cite{MPM}). The
crucial condition is therefore $E_{rung}<<E_{sph}$, and it is likely
to be satisfied (later, we will see that this is indeed the
case). Therefore, the applicability of the instanton--based approach
to high energy scattering is, at least {\it a priori}, plausible.

Later we will find it convenient to use the singular gauge for the
instanton field (the gauge in which the singularity of the potential
is shifted to the origin); the corresponding expression is
\be \label{insx}
A_\mu^a (x) \, = \, \frac{2\ \rho^2 {\overline \eta_{a \mu \nu}} 
x_\nu}{g \, x^2 (x^2 + \rho^2)}. 
\ee 
In momentum space, the instanton field is given by
\begin{equation}\label{insk}
A_\mu^a (k) \, = \, \frac{i (4 \pi)^2}{g} \, \frac{\overline{\eta}_{a
    \mu \nu} \, k_\nu}{(k^2)^2} \, \left[ 1 - \frac{1}{2} \, (k
  \rho)^2 \, K_2 (k \rho) \right].
\end{equation}

\section{Soft pomeron: the intercept}

In this section we will calculate the intercept of the soft pomeron
depicted in Fig. \ref{pom} to hadronic cross sections and will
calculate the intercept of the soft pomeron given by that equation.
We assume that the scale of the running coupling constant is being set
by the typical size of the QCD instanton $\rho_0$, and that the
coupling constant at that scale is small $\as (1/\rho_0) \, \ll \,
1$. This assumption is marginally well satisfied in QCD, but is
crucial for the use of QCD instantons. In the recent year there have
been proposed several scenarios of how the strong coupling constant
behaves at large distances \cite{dok,gri} suggesting that it never
becomes larger than $1$. If the ideas stated in \cite{dok,gri} are
correct then our small coupling assumption is likely to be
justified. Otherwise we note that the small coupling assumption is of
course better justified for electroweak interactions \cite{fut1}.

Assuming that $\as \, \ll \, 1$ we will neglect all the usual
perturbative QCD vertices as bringing extra powers of $\as$. We will
be resumming only the instanton--induced vertices, which give powers
of $\exp \left( - \frac{2 \pi}{\as} \right) \, \ll \, 1$ and our model
of the pomeron could be viewed as resumming powers of this parameter.

The structure of this section is the following. We will first derive
the intercept of the pomeron of Fig. \ref{pom}. Then we will estimate
the magnitude of virtual corrections to the real emission diagrams of
Fig. \ref{pom}.  We will then calculate the numerical value of the
pomeron intercept predicted by our model.

\subsection{Calculation of the intercept}

\begin{figure}
\begin{center}
\epsfxsize=5cm
\leavevmode
\hbox{ \epsffile{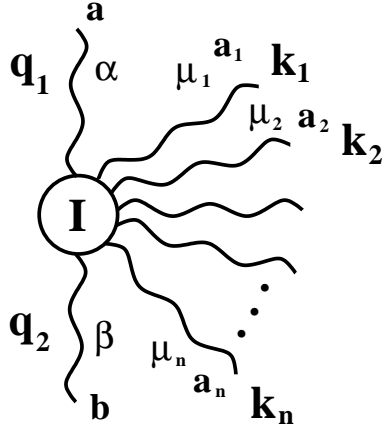}}
\end{center}
\caption{The multiple gluon vertex essential for construction of soft
  pomeron as described in the text.}
\label{gr}
\end{figure}

To be able to calculate the diagrams in Fig. \ref{pom} we have to
construct the basic vertex employed there --- the vertex coupling
several gluons to each other through an instanton. This will be done
in a way similar to the calculations of the baryon number violating
amplitudes involving instantons in electroweak theory done in
\cite{ring}. To calculate the multiple gluon vertex we have to
construct the Green function
\begin{equation}\label{grf1}
G^{a b a_1 \ldots a_n}_{\alpha \beta \mu_1 \ldots \mu_n} (q_1, q_2; 
k_1, \ldots , k_n) \, = \, (2 \pi)^4 \delta^4 (q_1 + q_2 - k_1 -
\ldots - k_n) \, \left< \, A_\alpha^a (q_1) \, A_\beta^b (q_2) \,
A_{\mu_1}^{a_1} (k_1) \, \ldots \, A_{\mu_n}^{a_n} (k_n) \, \right>,
\end{equation}
where $A_\mu^a (k)$ are instanton fields given by \eq{insk} and the
brackets $\left< \ldots \right>$ imply averaging over the instanton
sizes in the single instanton sector.  The Green function of Eq.
(\ref{grf1}) is depicted in Fig. \ref{gr}.  It combines two space-like
t-channel gluons carrying momenta $q_1$ and $q_2$ with $n$ final state
gluons with momenta $k_1 \, \ldots \, k_n$, which are taken to be on
the mass shell.  Thus, in Minkowski space, $q_i^2 = - {\underline
q}_i^2$, where ${\underline q}_i$ is the transverse component of the
gluon's momentum, and $k_i^2 = 0$. We need the Green function to be in
the momentum space representation because we are going to use it in
the momentum space diagram calculations below.  The difference between
our Green function and the one usually considered in the calculations
of $2 \rightarrow n$ process in QCD and electroweak theory
\cite{ring,bb,zakh} is that the two on-mass shell gluons in the
initial state of the electroweak process are substituted here by two
t-channel gluons which are far off mass shell.

Of course the Green function of \eq{grf1} describes the multiple gluon
interactions through an instanton only at not very high energies. As
energy increases the quantum corrections start becoming important
\cite{mu}, as was discussed in the Introduction. At very high energies
of the incoming pair of gluons the quantum corrections modify the
behavior of the Green function and the corresponding $2 \rightarrow n$
cross section. In our approach we assume that at high center of mass
energies of the hadron--hadron (or quarkonium--quarkonium) scattering
the dominant contribution comes from the pomeron diagrams in
Fig. \ref{pom} with a large number of rungs in the ladder. Thus the
energy per rung, or, equivalently, per vertex in the ladder remains
not too large, justifying the use of the Green function of
\eq{grf1}. We will return to quantify this issue below.

For a gluon on the mass shell the instanton field of \eq{insk} becomes
\be\label{inms} A_\mu^a (k) \, \rightarrow \, \frac{i 4 \pi^2
  \rho^2}{g} \, \frac{\overline{\eta}_{a \mu \nu} \, k_\nu}{k^2}
\hspace*{1cm} \mbox{as} \hspace*{1cm} k^2 \rightarrow 0.  \ee
Substituting the field of \eq{inms} for each $k_i$ line into \eq{grf1}
and employing the full instanton field of \eq{insk} for the $q$ lines
we obtain 
\ben G^{a b a_1 \ldots a_n}_{\alpha \beta \mu_1 \ldots
  \mu_n} (q_1, q_2; k_1, \ldots , k_n) \, = \, (2 \pi)^4 \delta^4 (q_1
+ q_2 - k_1 - \ldots - k_n) \, \int_0^\infty \, d \rho \, n (\rho) \,
\rho^{2 n} \, \left( \frac{i 4 \pi^2}{g} \right)^n \, \frac{- (4
  \pi)^4}{g^2} 
\een 
\be\label{grf2} 
\times \left( \prod_{i=1}^n \,
  \frac{{\overline \eta}_{a_i \mu_i \nu_i} k_{\nu_i}}{k_i^2} \right)
\, \frac{\overline{\eta}_{a \alpha \gamma} \, q_{1 \gamma}}{(q_1^2)^2}
\, \left[ 1 - \frac{1}{2} \, (q_1 \rho)^2 \, K_2 (q_1 \rho) \right] \,
\frac{\overline{\eta}_{b \beta \delta} \, q_{2 \delta}}{(q_2^2)^2} \,
\left[ 1 - \frac{1}{2} \, (q_2 \rho)^2 \, K_2 (q_2 \rho) \right],  
\ee
where $n (\rho)$ is the size distribution of instantons. 
We want to obtain an effective vertex for multiple pomeron
interactions. For that we have to truncate the external legs of the
Green function of \eq{grf2}. To amputate the propagators of the
external gluon lines one has to also remove the numerators of these
propagators. For that we need to know the gauge condition of the
fields of \eq{insx} and \eq{insk}. Usually the field of \eq{insx} is
derived using the Schwinger gauge condition $x_\mu \, A_\mu \, = \,
0$. However, we note that it also satisfies the covariant gauge
condition $\partial_\mu \, A_\mu \, = \, 0$. Therefore we will be
using it as a covariant (Feynman) gauge field and will employ it in
the covariant gauge calculation of diagrams. Therefore, to truncate
the Green function of \eq{grf2} we have to just multiply it by $q_1^2
\, q_2^2 \, k_1^2 \, \ldots \, k_n^2$. The resulting amputated Green
function is
\ben \Gamma^{a b a_1 \ldots a_n}_{\alpha \beta \mu_1 \ldots
  \mu_n} (q_1, q_2; k_1, \ldots , k_n) \, = \, (2 \pi)^4 \delta^4 (q_1
+ q_2 - k_1 - \ldots - k_n) \, \int_0^\infty \, d \rho \, n (\rho) \,
\rho^{2 n} \, \left( \frac{i 4 \pi^2}{g} \right)^n \, \frac{- (4
  \pi)^4}{g^2} 
\een 
\be\label{amf} 
\times \left( \prod_{i=1}^n \,
  {\overline \eta}_{a_i \mu_i \nu_i} k_{\nu_i} \right)
\, \frac{\overline{\eta}_{a \alpha \gamma} \, q_{1 \gamma}}{q_1^2}
\, \left[ 1 - \frac{1}{2} \, (q_1 \rho)^2 \, K_2 (q_1 \rho) \right] \,
\frac{\overline{\eta}_{b \beta \delta} \, q_{2 \delta}}{q_2^2} \,
\left[ 1 - \frac{1}{2} \, (q_2 \rho)^2 \, K_2 (q_2 \rho) \right].  
\ee

Now we are in the position to calculate the intercept of the soft
pomeron in Fig. \ref{pom}. The calculation will be done in the
Minkowski space.  To perform it we have to first multiply the
amputated Green function of \eq{amf} by the numerators of the
propagators of the t-channel gluons in the light cone gauge
(Weizs\"{a}cker--Williams approximation)
\ben
\frac{q_{1 \alpha}^\perp}{q_{1 +}} \, \,
\frac{q_{2 \beta}^\perp}{q_{2 -}}, 
\een
where the gluon $q_1$ carries a large $+$ component of the light cone
momentum and the gluon $q_2$ carries a large $-$ component. This is a
standard procedure, which is described in \cite{glr} in greater
detail. Then we should multiply the gluon line factors $k_i$'s in
\eq{amf} by the polarization vectors $\epsilon_i^{\lambda_i} (k_i)$
and square the amputated Green function of \eq{amf} excluding the
$\delta$ function.  We should keep in mind that the color space
orientation of the instanton in the complex conjugate amplitude is, in
principle, different from the color space orientation of the instanton
in the amplitude \cite{tin,son,bb,bb2,zakh}. Therefore we will have to
average over all possible orientations, for which we will be using the
approximation introduced in \cite{zakh}. We will integrate the
obtained expression over the phase space of the produced gluons and
sum over colors and polarizations $\lambda_i$
\ben
\frac{1}{n !} \, \prod_{i=1}^n \, \frac{d^3 k_i}{2 (2 \pi)^3 \omega_i}
\, \sum_{\lambda_i}
\een
where $\omega_i \, = \, k^i_4$ and sum over $n$ runs from $1$ to
infinity. We also have to multiply everything by one of each of the
denominators of the propagators of the t-channel gluons \cite{glr}
\ben
\frac{1}{{\underline q}_1^2} \, \,
\frac{1}{{\underline q}_2^2}
\een
and average over the colors of gluons $q_1$ and $q_2$, which would
give a factor of $1/(N_c^2 - 1)^2$. We also have to include the
symmetry factor of $(n !)^2$. We end up with the following expression
for the kernel of the integral equation describing the pomeron of
Fig. \ref{pom}
\ben
K (q_1^\perp, q_2^\perp) \, \ln \frac{1}{x} \, = \, \frac{1}{(N_c^2 -
1)^2} \,
\sum_{n=1}^\infty \, \int \, d^4 q
\, \delta^4 (q - \sum_{i=1}^n k_i) \, n !  
\een
\ben
\times \left( \prod_{i=1}^n
\, \frac{d^3 k_i}{2 (2 \pi)^3 \omega_i} \, \int \, d \xi \, \sum_{\lambda_i} 
\, \left[ {\overline \eta}_{a_i \mu_i \nu_i} k_{\nu_i} \,
\epsilon_{\mu_i}^{\lambda_i} (k_i) \right] \, 
\left[ U^{a_i b_i} (\xi) \, 
{\overline \eta}_{b_i \mu_i \nu_i} k_{\nu_i} \,
\epsilon_{\mu_i}^{\lambda_i} (k_i) \right]^*  \right) 
\een
\ben
\times \left( \frac{4
    \pi^2}{g} \right)^{2 n} \, \frac{(4 \pi)^8}{g^4}
\, \frac{1}{{\underline q}_1^2 \, {\underline q}_2^2} \, \left[
  \frac{{\overline \eta}_{a \alpha -} \, q_{1 \alpha}^\perp \,
    {\overline \eta}_{b \beta +} q_{2 \beta}^\perp  }{{\underline
      q}_1^2 \, {\underline q}_2^2} \right] \,  \left[
  \frac{U^{a a'} (\xi) \, {\overline \eta}_{a' \alpha -} \, 
  q_{1 \alpha}^\perp \,  U^{b b'} (\xi) \, {\overline \eta}_{b' \beta +} 
  q_{2 \beta}^\perp  }{{\underline q}_1^2 \, {\underline q}_2^2} \right]^* 
\een
\be\label{int1}
\times \, \left[ 
\int_0^\infty \, d \rho \, n (\rho) \, \rho^{2 n} \left( 1 - \frac{1}{2} 
\, (q_1^\perp
    \rho)^2 \, K_2 (q_1^\perp \rho) \right) \, \left( 1 - \frac{1}{2} 
\, (q_2^\perp \rho)^2 \, K_2 (q_2^\perp \rho) \right)  \right]^2,
\ee 
with $q \, = \, q_1 \, + \, q_2$. $U^{a b} (\xi)$ is the matrix of
global $SU(2)$ rotations in the adjoint representation and it is
responsible for rotating the instanton in the complex conjugate
amplitude in the color space \cite{tin,zakh}. We can perform the
summation over polarizations and averaging over the color space
orientations of the instantons using the approximation outlined in
\cite{zakh} 
\be\label{gl}
\int \, d \xi \, \prod_{i=1}^n \, \sum_{\lambda_i} \, \left[ {\overline 
\eta}_{a_i \mu_i \nu_i} \, k_{\nu_i} \, \epsilon_{\mu_i}^{\lambda_i} 
(k_i) \right] \, \left[ U^{a_i b_i} (\xi) {\overline
\eta}_{b_i \mu_i \nu_i} \, k_{\nu_i} \, \epsilon_{\mu_i}^{\lambda_i} 
(k_i) \right]^* \, \approx \, \left( \frac{E_{sph}^2}{3 n^2} \right)^n \, \approx \,
\frac{2 \, \pi \, n \, E_{sph}^{2 n}}{3^n \, e^{2 n} \, (n !)^2} 
\ee
for large n. \eq{gl} can be understood in the rest frame of the system
of two t-channel gluons. There the dominant contribution to \eq{gl}
comes from equally distributing center of mass energy among $n$ final
state gluons, which leads to the $(E^2_{sph}/n^2)^n$ in \eq{gl}.  The
estimate of \eq{gl} is somewhat crude, but was justified in
\cite{bb,bb2,zakh}. In writing \eq{gl} in the approximation outlined in 
\cite{zakh} we are likely to omit factors of $n$. However, as could be seen 
from the calculations below, we do not have the precision to keep all
the powers of $n$ at this point. Extra powers of $n$ may arise from
different ways of averaging over the instanton sizes (see \eq{peak}
and \eq{peak2}). The $q^2$ integration in \eq{int1} is sharply peaked
at the upper cutoff, as will be seen from the calculations presented
below, which we will put to be $E_{sph}$. This allowed us to just put
$E_{sph}^2$ in \eq{gl}. However, the uncertainty in our knowledge of
$E_{sph}^2$ may also introduce significant numerical changes, which we
can not control at the moment. We are going to clarify those in the
subsequent work \cite{fut2}.  We have also put $U^{ab} (\xi) \, =
\, 1$ for the t-channel gluon lines. As the left hand side of \eq{gl}
should of course be a Lorentz invariant expression in Minkowski space
it can only depend on boost invariant quantities. Thus we can see that
for $n=1$ the answer should be proportional to $k^2 = 0$ and should,
therefore, be zero. For higher values of $n$ the left hand side of
\eq{gl} may depend on the products of momenta of different gluons,
which results in a non-zero answer.

In evaluating the expression in \eq{int1} we have to note that the
perturbative size distribution of the QCD instantons $n (\rho)$ is
divergent at large sizes $\rho$. However, the lattice data
\cite{rslat} suggests that the distribution actually starts to fall off very
steeply with increasing $\rho$ after some critical value of $\rho \, =
\, \rho_0$. This behavior could be expected by the following argument based 
on the low energy theorem of \eq{lowen}, which has become known as
``$b/4$'' problem \cite{dd,et,svz,dia}. The issue is the following: it
is impossible to satisfy the low energy theorem of \eq{lowen} with a
purely classical field. The energy-momentum tensor of a classical
field given by \eq{emt} has only one power of $b$ in it, since the
classical field has no information about renormalization and running
coupling constant in it. Thus if one tries to satisfy \eq{lowen} with
this energy momentum tensor one would get an extra power of $b$ on the
left hand side which would not cancel. The commonly accepted
resolution of this problem is to assume that the instantons in the
instanton gas interact with each other, which leads to a different
infrared cutoff on the integrals over $\rho$ on the left and right
hand sides of \eq{lowen}
\cite{svz}. This leads us to conclude that QCD regularizes itself at 
large distances by cutting off the growth of the instanton size
distribution through some non-perturbative, but not classical
mechanism. This conclusion seems to be supported by lattice data
\cite{rslat}. For the purpose of this paper, in order to cure the 
problem we have to introduce an upper cutoff $\rho_0$ in the $\rho$
integrals in \eq{int1}.

In our approximation the sum over polarizations and colors (\eq{gl})
became independent of $k_i$'s. Therefore we can perform the
integration over the phase space of $n$ gluons. The integration yields
\be\label{keijo} 
\delta^4 (q - \sum_{i=1}^n k_i) \, \prod_{i=1}^n \, \frac{d^3 k_i}{2 
\omega_i} \, = \, \frac{(\pi/2)^{n-1} (q^2)^{n-2}}{(n-1)! \, (n-2)!}.  
\ee
After some simple algebra \eq{int1} becomes
\ben
K (q_1^\perp, q_2^\perp ) \, \ln \frac{1}{x} \, = \, \frac{1}{(N_c^2 - 1)^2} \, 
\sum_{n=2}^\infty \, \int \, d^2 {\underline q} \, \frac{d q_+}{2 \,
  q_+} \, d q^2 \,  n ! \, \frac{1}{(2 \pi)^{3 n}} \, 
\frac{(\pi/2)^{n-1} (q^2)^{n-2}}{(n-1)! \, (n-2)!} \, 
\frac{2 \, \pi \, n \, E_{sph}^{2 n}}{3^n \, e^{2 n} \, (n !)^2} \, \left( \frac{4
    \pi^2}{g} \right)^{2 n}
\een
\be\label{int2}
\times  \frac{(4 \pi)^8}{g^4} \,
\frac{1}{q_{1 \perp}^{4} \, q_{2 \perp}^{4} } \,  \left[ 
\int_0^{\rho_0} \, d \rho \, n (\rho) \, \rho^{2 n} \, 
\left( 1 - \frac{1}{2} \, (q_1^\perp
    \rho)^2 \, K_2 (q_1^\perp \rho) \right) \, \left( 1 - \frac{1}{2}
    \, (q_2^\perp \rho)^2 \, K_2 (q_2^\perp \rho) \right) \right]^2 .
\ee
We note again that the instanton size distribution is very sharply
peaked around its' maximum given by $\rho_0$. This statement is
supported by the lattice data \cite{rslat} and by various instanton
models
\cite{shur}. The size distribution for $\rho \, < \, \rho_0$ is very
close to the perturbative results \cite{rslat,size}, which sharply
increases near $\rho_0$. For $\rho \, > \, \rho_0$ the distribution
falls off steeply \cite{rslat}, justifying our cutoff procedure in
\eq{int2}. Thus without loosing the accuracy of our calculations we can put 
$\rho \, = \, \rho_0$ in the factors containing the confluent Bessel
functions in \eq{int2}.  Now we can see that the transverse momentum
integration factorizes in the expression for the kernel of the
equation for the pomeron--mediated forward amplitude
(\ref{int2}). Each segment of the ladder in Fig. \ref{pom}, consisting
of the square of the Green function of Fig. \ref{gr}, gives a factor
of
\be\label{tr1}
K(q^{\perp})\,\,\equiv\,\,\frac{1}{q_\perp^4} \, \left( 1 - \frac{1}{2} \,
(q_\perp
    \rho_0)^2 \, K_2 (q_\perp \rho_0) \right)^2 
\ee
to the rung of the ladder above it and a similar factor (with a
different momentum) to the rung below it. Therefore each rung of the
ladder brings in a square of the factor in \eq{tr1} integrated over
the transverse momentum
\be\label{i0}
I_0 \, \rho_0^6 \,  = \, \int \, d^2 q_\perp \,
\frac{1}{q_\perp^8} \, \left( 1 - \frac{1}{2} \, 
(q_\perp \rho_0)^2 \, K_2 (q_\perp \rho_0) \right)^4 .
\ee
Therefore, to obtain the pomeron's intercept out of the integral
kernel in \eq{int2} we should substitute the integral $I_0$ instead of
the transverse momentum integration in \eq{int2}. $I_0$ is easy to
evaluate numerically, which gives
\be\label{i01}
I_0 \, \approx \, 0.014 .
\ee
Integration over $q_+$ in \eq{int2} gives a factor of $\ln
\frac{1}{x}$.  Actually, the summation of the diagrams in Fig. 
\ref{pom} can be performed by the following simple integral equation
\be \label{eq1}
\phi(q^{\perp}_1, Y = \ln(1/x)) \,\,= \,\, K (q^{\perp}_1) \,\, +  
\,\,C K(q^{\perp}_1)\int^{Y}_0 \, d y  \,\int d^2 q^{\perp}_2 \, 
K(q^{\perp}_2) \, \phi(q^{\perp}_2, y)
\ee
where the kernel of \eq{int2} is $ K (q_1^\perp, q_2^\perp ) = C
K(q^{\perp}_1) \cdot K(q^{\perp}_2)$ and constant $C$ absorbs sum over
$n$ and integral over $q^2$ in \eq{int2}. $ \phi(q^{\perp}, Y =
\ln(1/x))$ denotes the sum of all diagrams of Fig. \ref{pom}--type, i. e., 
the pomeron--induced structure function or cross section. The initial
condition is given by $K(q^{\perp}_1)$, which is not crucial for us
and convenient for determination of the intercept. One can check that
\be
\phi(q^{\perp}, Y ) = K (q^{\perp}) \cdot \exp(
\Delta_{soft} \, Y)
\ee
is the solution of \eq{eq1} with
\be \label{eq2}
\Delta_{soft} \,\,=\,\,C \int \,\,d^2 q^{\perp} \,\,K^2(q^{\perp})\,\,.
\ee

A much more serious problem is posed by the integration over $q^2$ and
summation over $n$ ( both absorbed in $C$ in \eq{eq1} ) in
\eq{int2}, which is potentially very dangerous if one allows the upper
limit of the integration to be as large as the center of mass energy
of the system. This is what \eq{int2} seems to suggest if one takes it
at face value.  However, our \eq{int2} is valid only at relatively low
energies. As energy increases quantum corrections become important.
As was shown in \cite{br} the instanton-induced cross section for $2
\rightarrow n$ process in QCD falls off with energy for energies above
certain threshold energy $E_0$ due to quantum correction calculated in
\cite{mu}. Similar behavior results from the unitarity constraint
derived in \cite{zakh}. Here we assume that this is also the case for
our expression for the intercept. Namely we believe that at higher
energies per rung in the ladder the quantum corrections would
significantly slow down and later on completely overturn the growth of
the expression in \eq{int2} with energy, as was discussed in the
Introduction. These effects could be taken into account by putting an
effective upper cutoff $E_{sph}^2$ in the $q^2$ integration in
\eq{int2}. After performing all the integrations mentioned above
\eq{int2} yields the following expression for the pomeron's
intercept
\be\label{int3}
\Delta_{soft} \, = \, \frac{\pi \, I_0 \, \rho_0^4}{(N_c^2 - 1)^2} \, \frac{(4
  \pi)^6}{\alpha^3} \, \frac{E_{sph}^2 \, \rho_0^2}{6 \, e^2} \,
  \sum_{n=2}^\infty \, \frac{1}{[(n-1)!]^3}
\, \left( \frac{\pi \, E_{sph}^4}{6 \, \alpha \, e^2} \right)^{n-1}
\, \left( \int_0^{\rho_0} \, d \rho \, n (\rho) \, \rho^{2 n} \right)^2.
\ee
Employing the same argument about the sharpness of the distribution of
$n (\rho)$ \cite{rslat} as was used in performing the transverse
momentum integral in \eq{int2} we could rewrite
\be\label{peak}
\int_0^{\rho_0} \, d \rho \, n (\rho) \, \rho^{2 n} \, \approx \, \rho_0^{2
  n} \, N_0, 
\ee
where we have defined
\be
N_0 \, = \, \int_0^{\rho_0} \, d \rho \, n (\rho).
\ee
Substituting \eq{peak} into \eq{int3} we obtain
\be\label{int4}
\Delta_{soft} \, = \, \frac{\pi \, I_0 \, (N_0 \rho_0^4)^2}{(N_c^2 - 1)^2} 
\, \frac{(4 \pi)^6}{\alpha^3} \, \frac{E_{sph}^2 \, 
\rho_0^2}{6 \, e^2} \,
  \sum_{n=2}^\infty \, \frac{1}{[(n-1)!]^3}
\, \left( \frac{\pi \, E_{sph}^4 \, \rho_0^4}{6 \, \alpha \, e^2} \right)^{n-1}.
\ee
Note that $N_0$ has the dimensions of $M^{-4}$ and, therefore,
\eq{int4} is, of course, dimensionless. We can rewrite \eq{int4} in
a more compact form with the help of a generalized hypergeometric
function
\be\label{int5}
\Delta_{soft} \, = \, \frac{\pi \, I_0 \, (N_0 \rho_0^4)^2}{(N_c^2 - 1)^2} 
\, \frac{(4 \pi)^6}{\alpha^3} \, \frac{E_{sph}^2 \, 
\rho_0^2}{6 \, e^2} \, \left[ _0F_2 \left(-; 1, 1;  \frac{\pi \, E_{sph}^4 \, 
\rho_0^4}{6 \, \alpha \, e^2} \right) - 1 \right] .
\ee

One could be a little more careful in evaluating \eq{int3}. We can use
the perturbative expression for the size distribution of instantons
\cite{et,size}
\be\label{psize}
n (\rho) \, = \, d \, \frac{1}{\rho^5} \, \left( \frac{2 \pi}{\alpha}
\right)^{2 N_c} \, e^{- \frac{2 \pi}{\alpha}} \, \left(
  \frac{\rho}{\rho_0}  \right)^b,
\ee
where $d$ is given by 
\be
d \, = \, \frac{0.466 e^{-1.679 N_c}}{(N_c - 1)! (N_c - 2)!} \,
\approx \, 1.5 \, 10^{-3} \hspace*{1cm} \mbox{for} \hspace*{1cm}  N_c = 3 
\ee
and $b \, = \, (11/3) \, N_c$ for pure gluodynamics. We also used
$1/\rho_0$ as a renormalization scale in \eq{psize}. With the
distribution of \eq{psize} we obtain
\be\label{peak2}
\int_0^{\rho_0} \, d \rho \, n (\rho) \, \rho^{2 n} \, = \, d \, 
\left( \frac{2 \pi}{\alpha} \right)^{2 N_c} \, e^{- \frac{2
    \pi}{\alpha}} \, \frac{\rho_0^{2 n - 4}}{2 n + b - 4},
\ee
which, after plugging it into \eq{int3} and summation over $n$ in it
yields the following expression for the soft pomeron's intercept
\be\label{int7}
\Delta_{soft} \, = \, \frac{\pi \, I_0 \, d^2}{(N_c^2 - 1)^2} \, 
\left( \frac{2 \pi}{\alpha} \right)^{4 N_c} \, e^{- \frac{4
    \pi}{\alpha}} \, \frac{(4 \pi)^6}{\alpha^3} \, \frac{1}{81} \, 
\frac{E_{sph}^2 \, \rho_0^2}{6 \, e^2} \, \left[
_2F_4 \left(\frac{9}{2},\frac{9}{2} ; 1, 1, \frac{11}{2},
  \frac{11}{2}; \frac{\pi \, E_{sph}^4 \, \rho_0^4}{6 \, 
\alpha \, e^2} \right) - 1 \right],
\ee
where we have explicitly used $b = 11$ for $N_c = 3$. The perturbative
instanton size distribution of \eq{psize} has been shown to fit
very well the recent lattice data for $\rho \, < \, \rho_0$
\cite{rslat}. Therefore the estimate of the intercept given in
\eq{int7} is probably more precise than the estimate from
\eq{int5}. Though in both cases we have to make the same assumptions
to simplify the transverse momentum integration in \eq{int2}. We also
had to put a cutoff on $\rho$ integration in arriving to both Eqs. (\ref{int5})
and (\ref{int7}), which is not a very bad approximation judging from
lattice data \cite{rslat}, but still is an approximation.  In
Sect. IIIC we will compare numerical predictions of \eq{int5} and
\eq{int7} for the soft pomeron's intercept. 

 It should be stressed that the approach developed here in \eq{eq1} --
\eq{int7} is an example of the multiperipheral approach (see
Ref. \cite{MPM} and references therein). It gives all typical
properties of the multiperipheral ``ladder'' diagrams, namely, the
typical transverse momentum of produced particles (gluons) which does
not depend on the total energy ($ q^{\perp} \rho_0\, \,\approx 2
\div 3 $ in our case, which follows from \eq{tr1}) and the average 
energy $\hat s $ between the produced two bunches of gluons in
Fig. \ref{pom} which also does not depend on the total energy and is
equal to
\be \label{mp1}
  \Delta_{soft} \,\cdot\,\ln({\hat s} /E_{sph}^2) \, \sim \, 1 \,\,.
\ee
These features of our approach cannot be spoiled by virtual
corrections but, of course, they have to be taken into account when
one considers the numerical value of the pomeron intercept.

\subsection{Virtual corrections}

\begin{figure}
\begin{center}
\epsfxsize=10cm
\leavevmode
\hbox{ \epsffile{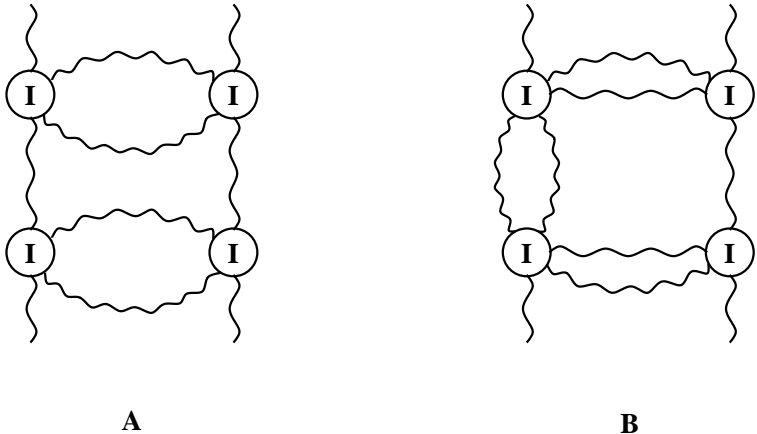}}
\end{center}
\caption{(A) A ``real'' diagram contributing to the pomeron intercept
  calculated in Sect. IIIA; (B) An example of a  virtual correction.}
\label{vt1}
\end{figure}

There are two types of virtual contributions to the soft pomeron we
are discussing. One is the additional gluon exchanges between
instantons in t-channel, an example of which is shown in Fig.
\ref{vt1}B. These corrections do not introduce extra logarithms of
energy ($\ln \frac{1}{x}$) in the problem. The other type of virtual
contributions could come from reggeization of the gluon propagator due
to the instanton interactions. This effects of course may introduce
additional powers of $\ln \frac{1}{x}$.

Let us first concentrate on the corrections brought in by the
t-channel gluons as demonstrated in Fig. \ref{vt1}B. We are going to
compare the contribution of the real diagram which is included in our
model of the soft pomeron above, given by Fig. \ref{vt1}A to the
contribution of the same diagram with a virtual correction added to
it, which is depicted in Fig. \ref{vt1}B.

\begin{figure}
\begin{center}
\epsfysize=5cm
\leavevmode
\hbox{ \epsffile{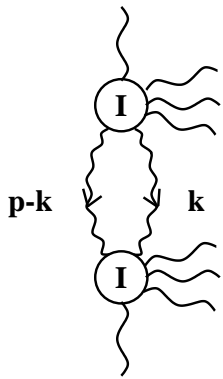}}
\end{center}
\caption{Virtual diagram of Fig. 4B, which contribution is 
calculated in the text.}
\label{vt2}
\end{figure}

The difference between the graphs in Figs. \ref{vt1}A and \ref{vt1}B
is just in extra gluon line, which introduces a loop integral. After
some simple algebra one can see that the integral brings in a factor
\be\label{loop}
K \, = \, \left( \frac{4 \pi^2}{g} \right)^2
\, \int \, \frac{d^4 k}{(2 \pi)^4}
\, \frac{\left[ 1 - \frac{1}{2} \, 
(k \rho_0)^2 \, K_2 (k \rho_0) \right]^2 \, \left[ 1 - \frac{1}{2} \, 
(|p - k| \rho_0)^2 \, K_2 (|p - k| \rho_0) \right]^2}{(k^2)^2 \, [(p - k)^2]^2},
\ee
where the labeling of the momenta is explained in Fig. \ref{vt2}. One
of the t-channel virtual lines carries the 4-momentum $k$ and the
other one carries $p-k$. In \eq{loop} we included only the factors
which make the contribution of the diagram in Fig. \ref{vt1}B
different from the contribution of the graph in Fig. \ref{vt1}A. We
also assumed that the instantons in the amplitude and in the complex
conjugate amplitude have the same orientation in the color space. This
assumption is not going to change our final answer by a large factor,
and, therefore is good for the estimate we are going to perform here.

The value of the integral in \eq{loop} is easy to estimate. First we
note that the expression in \eq{loop} is a decreasing function of $p$.
Therefore one can obtain an upper bound on the value of the integral
(\ref{loop}) by estimating it numerically for $p = 0$. The result
reads
\be\label{k}
K \, = \, 0.0052 \, \frac{\pi}{2 \, \alpha} \, \rho_0^4 .
\ee
We have to compare the factor given by \eq{k} to the factor given by
the $p$-line in Fig. \ref{vt1}A, which is just $1/(p^2)^2$. For the
estimate we can assume that approximately $p \, \approx \, 1/\rho_0$
(see \eq{tr1}). Therefore we conclude that the diagram in
Fig. \ref{vt1}B is different from the diagram in Fig. \ref{vt1}A by
the factor of $0.0052 \, \pi / 2 \, \alpha$. For the values of
$\alpha$ taken at the scale corresponding to the instanton size
($\alpha (1/\rho_0) \, \approx \, 0.6$) this number is still very
small, of the order of $0.01$. We can see that resummation of
t-channel virtual corrections, which are not enhanced by logarithms of
energy can not give us a large contribution and could be neglected.

One might think that the diagram of Fig. \ref{vt2} taken without the
s-channel gluon lines would give us a contribution to the mass of a
gluon, thereby creating a non-zero gluon's mass. However, in
constructing \eq{loop} we did not impose the condition that the gluons
forming the gluon loop should be in the color octet state, as one
should do in calculating the correction for the gluon's mass. That
condition was not used since with the s-channel gluons present in
Fig. \ref{vt2} the color of the pair of t-channel gluons is not
fixed. Now we have to point out that for $p = 0$ the contribution of
this diagram without the s-channel gluons is zero, which could be seen
after projecting the t-channel gluons onto the color octet state.

The second type of virtual corrections are enhanced by the factors of
$\ln \frac{1}{x}$ and are similar to gluon reggeization corrections
for the BFKL pomeron \cite{EAK,Yay}. To include these corrections
which bring in the evolution in $\ln \frac{1}{x}$ one has to calculate
the intercept of the same pomeron ladder of the type shown in
Fig. \ref{pom} with two t-channel gluons forming a color octet
combination and with the non-zero momentum transfer in the ladder. The
two gluons in the t-channel should in turn include all the virtual
corrections in them. A careful inclusion of the gluon reggeization
proved to be a very complicated task and will be completed elsewhere
\cite{fut2}. here we are going to estimate these corrections. 

Let us calculate the contribution of a single rung in the ladder of
Fig. \ref{pom}, with the ladder taken in the octet state with non-zero
momentum transfer $t = - {\underline q}^2$ (see also
Fig. \ref{slo}). A direct repetition of calculations in section IIIA
leads to the following expression for the octet channel integral $I_8
(q^2)$ which replaces integral $I_0$ in \eq{int5} and \eq{int7} for
the intercept of the Pomeron.
\ben
I_{8} (q^2) \, \rho_0^6 \, = \, \int \, d^2 k_\perp \,
\frac{|\left({\underline k} - \frac{1}{2} {\underline
    q}\right) \times \left({\underline k} + \frac{1}{2} {\underline
    q}\right) |^2}{[({\underline k} + \frac{1}{2}
  {\underline q})^2]^3 \, [({\underline k} - \frac{1}{2} {\underline
    q})^2]^3} \, \left[ 1 - \frac{1}{2} \, 
\left({\underline k} + \frac{1}{2} {\underline q}\right)^2 \rho_0^2 \, K_2
\left(|{\underline k} + \frac{1}{2} {\underline q}| \rho_0\right) \right]^2 
\een
\be \label{gr1}
\times \, \left[ 1 - \frac{1}{2} \, \left({\underline k} - \frac{1}{2}
 {\underline
    q}\right)^2 \rho_0^2 \, K_2 \left(|{\underline k} - \frac{1}{2}
 {\underline
    q}| \rho_0\right) \right]^2 ,
\ee
We will now assume that the virtual corrections to our pomeron lead to
reggeization of the t-channel gluons. Moreover, to estimate the
trajectory we will assume that the expression for the gluon Regge
trajectory has a form
\be \label{gr2}
\alpha_G (q^2) \,\,=\,\, 1\,\,+\,\,\Delta_{soft} \,\frac{I_{8}
(q^2)}{I_0}\,\,.
\ee
\eq{gr2} is our estimate of the gluon's trajectory. One can see that
$I_8 (q^2) \rightarrow 0$ at $q^2 \rightarrow 0$.

To simplify the expression for the gluon's Regge trajectory we note
that approximately
\be\label{appr}
\frac{1}{z^4} \, \left[1 - \frac{1}{2} \, z^2 \, K_2 (z)\right]^2 \, 
\approx \, \frac{e^{- z^2/2}}{16}.
\ee
Using \eq{appr} we can easily calculate function $I_8 (q^2)$, namely,
it turns out that
\be \label{gt2}
I_8 (q^2)\,\,=\,\, \frac{\pi}{256} \,\,\left\{\, 2 \, \left( \, e^{-
\frac{{\underline q}^2 \rho_0^2}{4}}\,\,-\,\,
 e^{- \frac{{\underline q}^2 \rho_0^2}{2}}\,\right)
\,\,+\,\,\frac{{\underline q}^2 \rho_0^2}{2}\,\left[\,Ei\left( - 
\frac{{\underline q}^2 \rho_0^2}{4}
\right)\,\, - \,\, 2 \, Ei\left( 
- \frac{{\underline q}^2 \rho_0^2}{2}\right)\,\right] \, \right\} \, \, .
\ee
\eq{gt2} should be substituted into \eq{gr2} to obtain our approximation 
for the gluon's Regge trajectory.

Now we have to determine whether the reggeization corrections can
significantly influence the value of the our pomeron's intercept. It
turns out that gluon reggeization can change the value of
$\Delta_{soft}$ since it leads to an additional factor in
\eq{i0}, which reduces this equation to the form
\be\label{ig0}
I_0 \, \rho_0^6 \, = \, \int \, d^2 q_\perp \,
\frac{1}{q_\perp^8} \, \left( 1 - \frac{1}{2} \,
(q_\perp \rho_0)^2 \, K_2 (q_\perp \rho_0) \right)^4 \,\,\cdot\,\,e^{ -
2\,[\alpha_G( q^2_{\perp}) - 1]\,\ln(\hat s/E_{sph}^2)}\,\, ,
\ee
where $\hat s$ is the energy in one rung of the ladder (see above).
As we have discussed the dominant contribution in the integral of
\eq{i0} comes from $q^2 \rho^2_0 \approx 2 - 3$ (see \eq{appr}) where
$2\,[ \alpha_G( q^2_{\perp}= 2/\rho^2_0) - 1]\,\ln(\hat
s/E_{sph}^2)\,\,\approx 0.4 \div 0.5 $, since $\ln(\hat s/E_{sph}^2)
\,\,\sim \,\,1/\Delta_{soft}$ ( see \eq{mp1} ).  Therefore, the additional
form factor in \eq{ig0} due to gluon reggeization can suppress the
value of $I_0$, and, consequently, the value of $\Delta_{soft}$ by a
factor of 2 $\div$ 3. Explicit numerical calculations confirm these
expectations. A more careful analysis of the contribution of virtual
corrections will be done later \cite{fut2}. In the numerical estimates
which will be done below we have to keep in mind that virtual
corrections can introduce a factor of 2 $\div$ 3 uncertainty in the
value of the pomeron's intercept.

\subsection{Value of the intercept}

The main purpose of this paper is to provide a qualitative mechanism
which could account for the effect of the soft pomeron. This was done
in Sect. IIIA. Nevertheless, we could try to estimate the numerical
value of the soft pomeron intercept obtained in Eqs. (\ref{int5}) and
(\ref{int7}) and check how it compares with the experimental value of
$0.08$ \cite{dl}.

\begin{figure}
\begin{center}
\epsfysize=5cm
\leavevmode
\hbox{ \epsffile{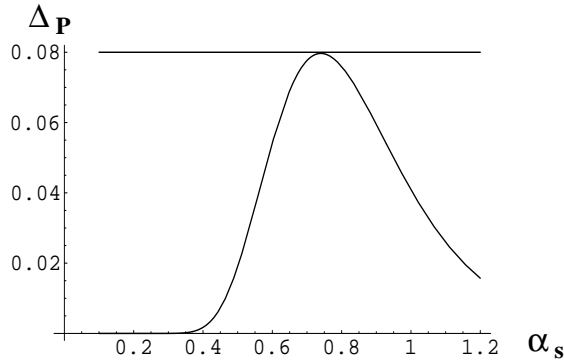}}
\end{center}
\caption{Intercept of our soft pomeron as a function of $\as$ for $E_{sph} 
= 2.4 \, \mbox{GeV}$, $\rho_0 = 0.3 fm$ and the virtual corrections giving a
factor of $\delta = 0.31$ suppression. The horizontal line corresponds
to the phenomenological pomeron intercept \protect\cite{dl}. }
\label{int}
\end{figure}

At the same time one has to be very careful with the numerical
estimates of the intercept in \eq{int5} and \eq{int7}. It is important
to realize that the intercept of \eq{int5} strongly depends on the
position of the maximum of the instanton size distribution $\rho_0$
and on the value of the integral of this distribution over all
$\rho$'s given by $N_0$. The value of the intercept in \eq{int7}
depends on the parameter $\rho_0$ since it determines the scale of the
strong coupling constant. The current knowledge of these parameters is
poor \cite{rslat,br,neg,sh}. The value of $N_0$ seems to be
undetermined up to an order of magnitude \cite{neg}.  Uncertainty in
the value of $\rho_0$ brings in an uncertainty in the value of the
strong coupling constant $\alpha (1/\rho_0)$.  Finally a very
important question concerns the value of the cutoff, which was taken
to be $E_{sph}$ in Eqs. (\ref{int5}) and (\ref{int7}). The cutoff
corresponds the maximum invariant mass of the particles produced in
each vertex. In electroweak physics it correspond to the energies when
the growth of the cross section of the $2 \rightarrow n$
instanton-induced process slows down and, maybe, even starts
decaying. In QCD it may correspond to the energy at which the
unitarity limit of the $2 \rightarrow n$ process is reached
\cite{zakh}.  The exact value of the energy at which this happens is
still not determined precisely \cite{mu,tin,son,br,zakh}. It is one of
the most crucial assumptions of this work that the turnover of the
cross section does happen in QCD and that the corresponding invariant
mass $E_{sph}^2$ is not too large.

The intercept of the pomeron calculated above (\eq{int7}) has been
derived for pure gluodynamics. If we want to compare our result to the
experimentally measured one \cite{dl} we have to include quarks in the
theory. They would introduce suppression of the amplitudes through the
Faddeev--Popov determinant \cite{th}. Quark lines could be included in
the scattering in two ways: they could be produced from an instanton
into the final state, in which case the suppression factor they
introduce is small, of the order of $10^{-5}$, so this type of quark
effects could be neglected. Alternatively the quarks could be absorbed
in the gluon condensate \cite{svz2}. In the latter case the
suppression factor in the distribution of instantons
\eq{psize} is
\cite{svz2}
\be\label{sup}
\prod_{q = u, d, s, \ldots } \, 1.3 \, \left(m_q \, \rho - \frac{2 \pi^2}{3} \, (0| 
{\overline q} q |0) \, \rho^3\right).
\ee
The suppression for the intercept of \eq{int7} is given by the square
of the factor in \eq{sup}. For our estimate we will just put $\rho \,
= \, \rho_0$ in \eq{sup} and use $(0| {\overline q} q |0) \, = \, -
(0.25 \, \mbox{GeV})^3$. Taking the product over three flavors we
obtain a suppression factor of $0.023$ for the intercept.

Keeping in mind all of the above mentioned limitations we substitute
the value of $E_{sph} \, = \, 2.4 \, \mbox{GeV}$ and $\rho_0 \, = \,
0.3 fm$ from \cite{sh} into \eq{int7} and plot the resulting value of
the pomeron's intercept as a function of the coupling constant
$\as$. We also included suppression introduced by virtual corrections
by including an extra factor $\delta$ on the right hand side of
\eq{int7}. The resulting pomeron's intercept is related to the intercept 
of \eq{int7} by the following relation
\be
\Delta_P \, = \, \delta \, \left[ \prod_{q = u, d, s, \ldots } \, 1.3 \, 
\left( m_q \, \rho_0 - \frac{2 \pi^2}{3} \, (0| 
{\overline q} q |0) \, \rho_0^3 \right)\right]^2 \, \Delta_{soft}
\ee
In the plot shown in Fig. \ref{int} we have put $\delta = 0.31$.  Of
course we have to admit that the plot in Fig. \ref{int} depends on the
value of the suppression factor $\delta$. It is also very sensitive to
the cutoff $E_{sph}$. A work on putting some constraints on these
parameters is under way \cite{fut2}. With the progress in the field of
instanton models the exact numerical value of the soft pomeron
intercept might be calculated with more confidence and precision.

The maximum of the intercept depicted in Fig. \ref{int} is close to
the experimental result of $0.08$, shown there by a horizontal
line. The value of the coupling constant at the scale set by the
instanton size $\rho_0$ is near the point on the plot of Fig. \ref{int}
where the maximum is achieved. Thus our value of the intercept should
be about $0.08$. Also, the pomeron's intercept seems to vanish at
small distances (large momenta), in agreement with experimental data
\cite{dl}. Also our model of soft pomeron allows a non-singular
continuation of the value of the intercept into the region of large
distances and large coupling constant. Even if the strong coupling
constant becomes very large at small momenta (large distances) the
intercept of our pomeron still remains small (see
Fig. \ref{int}). That way the strong interactions are able to yield us
with a small pomeron's intercept even in the kinematic region where
they are very strong. Here we have to recall that strictly speaking
our quasi--classical approach is applicable only when the coupling
constant is small. Nevertheless, the smallness of the intercept at
large coupling is an interesting result, supported by phenomenological
observations \cite{dl}.

\section{Soft pomeron: the slope and trajectory}

In this Section we are going to calculate the slope and the intercept
of the soft pomeron conjectured above.  We will include contribution
of the virtual corrections by putting a factor $\delta$ in the
expression for pomeron's trajectory (see Sect. IIIB). The quark
contribution will also be treated similarly to Sect. IIIC. Therefore
the slope and trajectory of the soft pomeron which will be derived
below will result only from the real emissions part of its
kernel. This will correspond to summing the diagrams of the type shown
in Fig. \ref{pom}.

\begin{figure}
\begin{center}
\epsfysize=5cm
\leavevmode
\hbox{ \epsffile{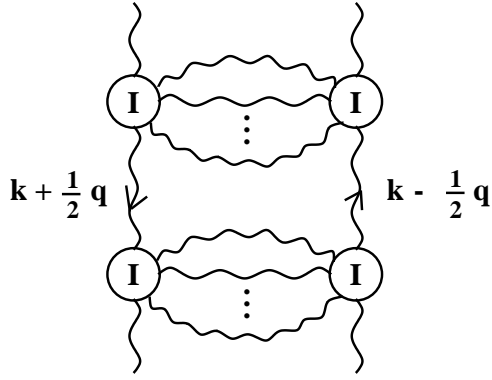}}
\end{center}
\caption{Soft pomeron with non-zero momentum transfer.}
\label{slo}
\end{figure}

To calculate the slope of the pomeron considered here we will follow
the standard procedure. Let us consider a pomeron with non-zero
momentum transfer described by Mandelstam variable $t$. A rung of the
off-forward pomeron's ladder is shown in Fig. \ref{slo}. One t-channel
gluon carries a momentum $k+ \frac{1}{2} q$, whereas the other one
carries momentum $k - \frac{1}{2} q$. Thus the momentum transfer is $t
\, = \, - {\underline q}^2$.

It is easy to see that the only difference between the pomeron ladder
with non-zero momentum transfer and the $t=0$ pomeron considered in
Sect. III is in the transverse momentum integral. An easy calculation
shows that for off-forward pomeron the integral $I_0$ in the
expression for the forward pomeron intercept (\ref{int7}) should be
replaced by
\ben
I (q^2) \, \rho_0^6 \, = \, \int \, d^2 k_\perp \,
\frac{({\underline k}^2 - \frac{1}{4} {\underline
    q}^2)^2}{[({\underline k} + \frac{1}{2}
  {\underline q})^2]^3 \, [({\underline k} - \frac{1}{2} {\underline
    q})^2]^3} \, \left[ 1 - \frac{1}{2} \, 
\left({\underline k} + \frac{1}{2} {\underline q}\right)^2 \rho_0^2 \, K_2
\left(|{\underline k} + \frac{1}{2} {\underline q}| \rho_0\right) \right]^2 
\een
\be\label{i}  
\times \, \left[ 1 - \frac{1}{2} \, \left({\underline k} - \frac{1}{2} {\underline
    q}\right)^2 \rho_0^2 \, K_2 \left(|{\underline k} - \frac{1}{2} {\underline
    q}| \rho_0\right) \right]^2 ,
\ee
which naturally reduces to \eq{i0} when ${\underline q} = 0$. To obtain the slope
of our pomeron we need to expand \eq{i} up to the quadratic term in
${\underline q}^2$. After expansion of \eq{i} in ${\underline q}^2$ and
integration over the angles we obtain
\be\label{i1}
I \, = \, I_0 \, + \, \pi \, {\underline q}^2 \, \rho_0^2 \, \int_0^\infty \,
d k \, \rho_0 \, \, L (k \rho_0)
\ee
where
\ben
L (k \rho_0) \, = \, - \, \frac{1}{32 (k \rho_0)^9} \, \left( 1 -
  \frac{1}{2} (k \rho_0)^2 K_2 (k
  \rho_0)\right)^2 \, \left[ 32 + (k \rho_0)^6 K_1 (k \rho_0)^2 + 4 (k
  \rho_0)^4 K_2 (k \rho_0) + 8 (k \rho_0)^4 K_2^2 (k \rho_0) \right. 
\een
\ben
- 2 (k \rho_0)^6
  K_2^2 (k \rho_0) + 2 \, (k \rho_0)^4 K_0 (k \rho_0) \left( 1
    - \frac{1}{2} (k \rho_0)^2  K_2 (k
  \rho_0)\right) - 20 (k \rho_0)^3 K_3 (k \rho_0) + 2 (k \rho_0)^5 K_2 (k
\rho_0) K_3 (k \rho_0)  
\een
\ben
+ (k \rho_0)^6 K_3^2 (k \rho_0) + 2 (k \rho_0)^3 K_1 (k
\rho_0) \left( -10 + (k \rho_0)^2 K_2 (k \rho_0) + (k \rho_0)^3 K_3 (k \rho_0)
\right) + 2 (k \rho_0)^4 K_4 (k \rho_0) 
\een
\be\label{l}
\left. - (k \rho_0)^6 K_2 (k \rho_0)
K_4 (k \rho_0) \right].
\ee
The integral in \eq{i1} is dominated by the small values of the
argument, as could be seen from the behavior of the function $L (k
\rho_0)$. We can expand the function $L (k \rho_0)$ of \eq{l} to obtain
\be\label{ll}
L (k \rho_0) \, \approx \, - \, \frac{1}{256 \, k \, \rho_0}
\ee
when the argument is small. The full integral of \eq{i} is convergent,
whereas the integral in \eq{i1} seems to be divergent in the
infrared. Therefore to regularize it after we plug \eq{ll} in it we
have to use $q$ as a lower cutoff and $1/\rho_0$ as an upper
cutoff. The result yields
\be\label{ii}
I \, = \, I_0 \, - \, \frac{\pi}{256} \, q^2 \, \rho_0^2 \, \ln
\frac{1}{q \rho_0}.
\ee 
Recalling that a pomeron trajectory is given in terms of its'
intercept and slope by
\be\label{reg}
\Delta (t) \, = \, \Delta \, + \, \alpha' \, t
\ee
and comparing (\ref{reg}) to \eq{int3} and \eq{ii} we can conclude
that for our pomeron
\be\label{a'}
\frac{\alpha'}{\Delta} \, = \, \frac{\pi \rho_0^2}{256 I_0} \, 
\ln \frac{1}{q \rho_0}.
\ee
Since there is still some $q$ dependence left in \eq{a'} we conclude
that our pomeron's trajectory is not quite linear in $t$ and the
notion of pomeron's slope is not very well defined for it. A similar
result was obtained for the soft pomeron trajectory in \cite{ga} using
Regge theory, which seemed to confirm experimental data of
\cite{hol}. The conclusion of \cite{ga} was drawn out of two-pion
exchange model of hadronic interactions.  We can obtain a numerical
estimate of the pomeron's slope of \eq{a'} by assuming that $ \ln
\frac{1}{q \rho_0} \, \sim \, 1$. Substituting the values of $I_0$
from \eq{i01} and $\rho_0 = 0.3 \, fm = 1.5 \, GeV^{-1}$ into \eq{a'} we obtain
\be\label{slp}
\frac{\alpha'}{\Delta} \, \approx \, 2.0 \, GeV^{-2},
\ee
which is remarkably close to the experimental result for the soft
pomeron \cite{dl}
\be
\left( \frac{\alpha'}{\Delta} \right)_{exp} \, = \, \frac{0.25}{0.08} 
\, GeV^{-2} \, \approx \, 3.1 \, GeV^{-2}. 
\ee

In principle the integral of \eq{i} substituted in the \eq{int3} (or
either \eq{int5} or \eq{int7}) can give us the full trajectory of our
pomeron. The integral in \eq{i} could be estimated numerically for
positive values of ${\underline q}^2$, which corresponds to the
negative values of $ t = -{\underline q}^2$. However, for positive $t$
the integral in \eq{i} has a singularity and if taken literally would
be simply divergent. The resolution of this problem is the following:
one should first perform the integration in \eq{i} for negative $t$
and then analytically continue the results into the region of positive
$t$. This should be done analytically and the exact analytical
treatment of the integral in \eq{i} seems to be rather
complicated. Instead we will use an approximation, which preserves all
the main qualitative features features of the answer.

Using \eq{appr} in \eq{i} and performing the integration over $k$
yields the following expression
\be\label{iii}
I (t) \, \approx \, \frac{\pi}{256} \, \left\{ 2 e^{t
\rho_0^2/2} - e^{t \rho_0^2/4} - \frac{t \rho_0^2}{2} \, \left[ 2 \,
Ei \left(\frac{t \rho_0^2}{2}\right) - Ei \left(\frac{t \rho_0^2}{4}\right)
\right] \right\}.
\ee
We can check the quality of the approximation done to obtain \eq{iii}
by comparing its' value for $t=0$ with $I_0$ in \eq{i01}. \eq{iii} gives
\be
I (0) \, \approx \,  \frac{\pi}{256} \, \approx \, 0.012,
\ee
which is very close to the exact value $I_0 = 0.014$ in \eq{i01}.  If
we expand \eq{iii} for small $t$ up to the terms linear in $t$ we
would exactly recover the expression for the slope of the trajectory
given by \eq{a'}! Thus \eq{iii} provides us with a very good approximation of 
the integral in \eq{i}. 

Our soft pomeron's trajectory could be obtained by substituting $I(t)$
from \eq{iii} into \eq{int7} instead of $I_0$. This gives
\ben
\Delta_{soft} (t) \, = \, \frac{\pi \, d^2}{(N_c^2 - 1)^2} \, 
\left( \frac{2 \pi}{\alpha} \right)^{4 N_c} \, e^{- \frac{4
    \pi}{\alpha}} \, \frac{(4 \pi)^6}{\alpha^3} \, \frac{1}{81} \,
    \frac{E_{sph}^2 \, \rho_0^2}{6 \, e^2} \, \left[ \, _2F_4
    \left(\frac{9}{2},\frac{9}{2} ; 1, 2, \frac{11}{2}, \frac{11}{2};
    \frac{\pi \, E_{sph}^4 \, \rho_0^4}{6 \, \alpha \, e^2} \right) -
    1 \right]
\een
\be\label{tra}
\times \frac{\pi}{256} \, \left\{ 2 e^{t
\rho_0^2/2} - e^{t \rho_0^2/4} - \frac{t \rho_0^2}{2} \, \left[ 2 \,
Ei \left(\frac{t \rho_0^2}{2}\right) - Ei \left(\frac{t \rho_0^2}{4}\right)
\right] \right\}.
\ee
After taking into consideration the virtual corrections and the quark
contributions the answer for the pomeron's trajectory becomes
\be\label{traf}
\Delta_P (t) \, = \, \delta \, \left[ \prod_{q = u, d, s, \ldots } \, 1.3 \, 
\left( m_q \, \rho_0 - \frac{2 \pi^2}{3} \, (0| 
{\overline q} q |0) \, \rho_0^3 \right)\right]^2 \, \Delta_{soft} (t),
\ee
where $\Delta_{soft} (t)$ is given by \eq{tra}

\begin{figure}
\begin{center}
\epsfxsize=10cm
\leavevmode
\hbox{ \epsffile{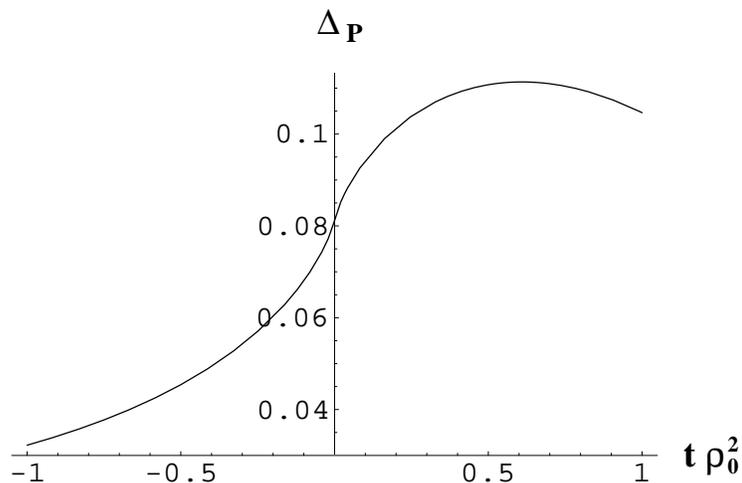}}
\end{center}
\caption{Soft pomeron's trajectory. $t$ is measured in the units 
of $\rho_0^2$.}
\label{traj}
\end{figure}

The soft pomeron's trajectory of \eq{traf} is depicted in
Fig. \ref{traj}. It is plotted for $\rho_0 = 0.3 fm$, $E_{sph} = 2.4
GeV$, $\delta = 0.31$ and $\alpha \approx 0.75$, i.e., the same values
as were used for the estimates of the pomeron's intercept in
Sect. IIIC.

As could be seen from Fig. \ref{traj} the soft pomeron's trajectory
never becomes negative for negative $t$, which means that the total
cross section generated by the soft pomeron exchange is always growing
with energy. For positive values of $t$ the trajectory first starts
growing but then slows down and even turns over and falls off. It does
not reproduce the linear growth behavior which would be predicted by
Regge theory. However, the linear behavior of the trajectory for
positive $t$ is usually associated with confinement in QCD. It is a
well known fact that the instanton--induced effects can not account
for QCD confinement \cite{cdg}. Therefore one should not expect that
the pomeron constructed out of instantons should exhibit such
confinement features as a linear trajectory for positive $t$. It is
most likely that different non-perturbative physical mechanisms play
an important r{\^o}le in that region.

\section{Summary and discussion}

In our approach we envision the soft pomeron as a chain of topological
transitions induced by high momentum gluons. This is the main point of
the paper. The corresponding pomeron ladder is depicted in
Fig. \ref{pom}. The multi--gluon vertices of the ladder are generated
through a strong quasi--classical vacuum field.

To quantify the pomeron illustrated in Fig. \ref{pom} we have used
the instanton fields, which have an explicit analytical form which we
could employ. The resulting expressions for the pomeron intercept are
given in Eqs. (\ref{int5}) and (\ref{int7}), corresponding to
different ways of averaging over the instanton sizes. We have
estimated the virtual corrections and showed that they are likely to
change the value of the soft pomeron's intercept by a factor of $2
\div 3$ at the most. The numerical value of the intercept has been
plotted in Fig. \ref{int} as a function of the running coupling
constant.  The resulting intercept at $\as (1/\rho_0) \, = \, 0.7$ is
very close to the phenomenological value of $0.08$ \cite{dl}. The
intercept falls off at small distances, which agrees with
phenomenological observations of \cite{dl}. At short distances (small
momenta) the dominant contribution to the structure functions comes
from the hard pomeron \cite{dl}. The work on understanding how our
soft pomeron merges with hard (BFKL) pomeron in that region is under
way \cite{fut3}.  At large distances (small momenta) the pomeron's
intercept also decreases. That way, in the region where the coupling
constant is large and the interactions are very strong the pomeron's
intercept can still be small. Thus our results could provide the
resolution of the long-standing puzzle of the smallness of the soft
pomeron intercept \cite{dl} in the strong coupling regime.

Finally, we have calculated the slope of the soft pomeron in
\eq{slp}. The value of the slope turned out to be close to the
phenomenologically suggested value of $0.25 \, \, \mbox{GeV}^{-2}$
\cite{dl}. The full trajectory of the pomeron has also been derived
(see \eq{tra} and \eq{traf}) and plotted in Fig. \ref{traj}. Since the
instantons can not account for confinement in QCD we do not expect a
linear trajectory for positive $t$. For negative $t$, as the absolute
value of $t$ is getting larger the trajectory ceases to be linear and
exhibits some curvature.

We can also estimate the multiplicity of the produced particles in a
hadron--hadron scattering event mediated by our pomeron of
Fig. \ref{pom}. The particles would be produced by multiperipheral
mechanism. From the estimate of \eq{mp1} we can conclude that the
density of instantons per unit of rapidity is rather small. However,
since there are several particles emitted coherently from each
instanton, we may expect correlations in particle production.

Let us estimate the typical number of particles produced on an
instanton in the ladder of Fig. \ref{pom}. For that purpose we will
consider the series of \eq{int4} and estimate the average $n$ in that
series. That would be a typical multiplicity of gluons produced off an
instanton. A simple calculation yields
\be\label{mult}
\left< n \right> \, = \, 1 + \frac{\pi \, E_{sph}^4 \, \rho_0^4}{6 \, \as e^2} \, \,
\frac{_0F_2 \left(-; 2,2; \frac{\pi E_{sph}^4 \, \rho_0^4}{6 \as e^2} \right)}
{_0F_2 \left(-; 1,1; \frac{\pi E_{sph}^4 \, \rho_0^4}{6 \as e^2} \right)}.
\ee
For the values of the parameter we have used above in estimating the
pomeron's intercept, $E_{sph} = 2.4 \, \mbox{GeV}$, $\rho_0 = 0.3 fm$,
$\as \approx 0.7$ (see Fig. \ref{int}), we obtain $\left< n \right>
\approx 3$. This result agrees with the estimate of \cite{kl}.

While our numerical results are very sensitive to a number of
instanton--related parameters, we believe that our approach can
provide an interesting new insight in the mechanism of high--energy
scattering and particle production in QCD.  Further theoretical and
experimental developments are clearly needed to clarify the
relationship between the soft pomeron and the properties of the QCD
vacuum.

\section*{Acknowledgments}

The authors are very much indebted to Ian Balitsky, Larry McLerran, Al
Mueller and Edward Shuryak for numerous helpful discussions about
instantons. We would like to thank James Bjorken, Greg Carter,
Hirotsugu Fujii, Asher Gotsman, Lev Lipatov, Uri Maor, Rob Pisarski,
Andreas Ringwald, Thomas Sch\"{a}fer, Fridger Schrempp, Dam Son,
Chung-I Tan, Arkady Vainshtein and Valentin Zakharov for many
informative and encouraging discussions. Yu. K. would like to thank
the High Energy Physics Department at Tel Aviv University for their
support and hospitality during the final stages of this
work. E. L. thanks BNL Nuclear Theory group for their hospitality and
creative atmosphere during several stages of this work.

This manuscript has been authored under Contract No. DE-AC02-98CH10886
with the U. S. Department of Energy. The research of Yu. K. and
E. L. was supported in part by the BSF grant $\#$ 9800276 and by
Israeli Science Foundation, founded by the Israeli Academy of Science
and Humanities.

\end{document}